# Exciton Binding Energy in small organic conjugated molecule.

Pabitra K. Nayak*

Department of Materials and Interfaces, Weizmann Institute of Science, Rehovot, 76100, Israel

Email: Pabitra.Nayak@weizmann.ac.il

**Abstract:** For small organic conjugated molecules the exciton binding energy can be calculated treating molecules as conductor, and is given by a simple relation BE ≈ $e^2/(4\pi\varepsilon_0\varepsilon R)$, where $\varepsilon$ is the dielectric constant and R is the equivalent radius of the molecule. However, if the molecule deviates from spherical shape, a minor correction factor should be added.

## 1. Introduction

An understanding of the energy levels in organic semiconductors is important for designing electronic devices and for understanding their function and performance. The absolute hole transport level and electron transport level can be obtained from theoretical calculations and also from photoemission experiments [1][2][3]. The difference between the two transport levels is referred to as the transport gap ($E_t$). The transport gap is different from the optical gap ($E_{opt}$) in organic semiconductors. This is because optical excitation gives rise to excitons rather than free carriers. These excitons are Frenkel excitons and localized to the molecule, hence an extra amount of energy, termed as exciton binding energy ($E_b$) is needed to produce free charge carriers. What is the magnitude of $E_b$ in organic semiconductors? Answer to this question is more relevant in emerging field of Organic Photovoltaic cells (OPV), due to their impact in determining the output voltage of the solar cells. A lot of efforts are also going on to develop new absorbing material for OPV. It is also important to guess the magnitude of exciton binding energy before the actual synthesis of materials.

Experimentally $E_b$ is determined by [3]

$$E_b = E_t - E_{opt} \quad (1)$$

Hill et al [3] used ultraviolet photoelectron spectroscopy (UPS) and inverse photoelectron spectroscopy (IPS) to get information about the transport gap. They identify the peaks in the UPS and IPS spectra with the Homo and Lumo levels and the transport gap with the difference between the Homo and Lumo levels. They estimated the exciton binding energy for a few molecules using eq.1. Whereas Krause et al [4] argue that the transport gap should be given by the onset of the peaks rather than the peak positions themselves and hence arrive at a smaller value of the exciton binding energy. Djurovich [5] et al have estimated the exciton binding energy for a large variety of molecules using data from different sources like photoemission, electrochemistry and DFT calculations to estimate the transport gap. The authors showed that as the band gap increases, the exciton binding energy also increases. Knupfer [6] presented a simple model to explain the exciton binding energy for many polymers and small molecules. Hummer et al [7] have calculated the exciton binding energy for oligoacenes and show that the exciton binding energy decreases as the size of the molecule increases. They attribute this to the increased delocalization of the carriers over the length of the molecule.

In this Letter, we present a simple (yet powerful) model to estimate $E_b$ of planar organic molecules. This model can also estimate $E_b$ with low-cost computational effort, without a prior knowledge of any experimental data.

**2. Computation Method.**

Density Functional Theory (DFT) was used in all the calculations using Gaussian03 [8]. The DFT method used here is based on the hybrid B3LYP functional [9]. The geometries of all the molecules were optimized at 6-31G (d) basis functions. Molecular cavity for Polarizable Continuum Model (PCM) was built by following the new definition using United Atom Model [10], i.e. by putting a sphere around each atom except Hydrogen. Hydrogen atoms are enclosed in the sphere of the atom to which they are bonded. Monte

Carlo integration procedure was used for the calculation of molecular volume. The volume calculation (at 6-31G (d) level) was repeated several times until the integrated density is close to the total number of electrons. Isotropic polarizability (at zero frequency) of molecules was calculated for the optimized geometry of molecules at 6-31G (d) level. Dielectric constant was calculated using Clausius- Mossetti equation. Radius of the molecule (to be used as solvent molecule) is defined as the radius of a sphere whose volume is equal to the molecular volume. Details of the calculations can be found in ref. [1][2]. Experimental optical gaps are determined from edges of aborsorbance spectra of thin films reported in literature.

## 3. Results and Discussion

Energy levels of molecules are modified in solid state due to interaction with molecules in the neighbourhood. Polarizable continuum model is shown to be successful in calculating energy level properties of molecules in condensed phase [11][12]. A method based on DFT/TD-DFT and polarizable continuum model (PCM) was described recently to determine energy levels in organic semiconductors [1][2]. In organic solids, molecules are solvated by the molecules of its own type. The solvent (here the solid matrix of molecules) is modeled as an isotropic, continuous dielectric that surrounds a solute (molecule of interest) cavity. This simple model is appropriate for amorphous and glassy solid films where the molecules are randomly oriented with respect to each other. Nevertheless it is first order of approximation to account for the dielectric effect on energy levels. In order to define parameters in PCM we used consistently the calculated values for dielectric constant and molecular volume, so that calculation can be done for any new molecule without reference to the experimental value (though experimental values are known for some molecules , see table 1).

Though the model overlooks the crystal packing of molecules in solid state, it certainly gives a quick estimation of energy levels. In particular this model is suitable for developing new molecules, as crystal packing in intermolecular interactions is unknown a priori.

Energy levels (ionization potential, electron affinity, optical gap, transport gap and binding energy) were calculated for 13 planar organic molecules which are shown in Table 1. We would like to point out here that, these calculated parameters shown nice correlation with experimentally determined energy levels by photoelectron spectroscopy [1][2]. We also put experimentally determined molecular Volume and optical gap of the molecules, in table 1 for a comparison to our calculated values (see ref [1][2] for other experimentally determined values) . For consistency (as the combined experimental errors are of few hundreds of meV) we use the calculated values for determining exciton binding energy.

*Exciton binding energy:*

We can think of the exciton binding energy as follows. Optical excitation in organic semiconductors is primarily excitonic. The bound hole and electron pair is confined to the molecule. This simple picture of understanding the exciton binding energy is taking out the electron form bound state (exciton) and putting onto another molecular unit far away [6]. Figure 1 schematically describes the whole scenario. In this process one has to pay the charging energy, (which is same as exciton binding energy) to make a negatively charged and a positively charged molecular conductor which are far away from each other. Evidences support that molecules behave similarly like macroscopic bodies, with the shape and dimensions influences the energies through their capacitance [13]. Hence $E_b$ can be written as

$E_b = 2\times$ Charging of Molecular conductor   (2)

Or $E_b = e^2/C$, where C is the capacitance of molecule    (3)

With principles based on simple electrostatics and shape and size of the molecule, the capacitance (C) of molecules can be calculated as [14]

$$C = f\varepsilon_0\varepsilon\sqrt{(4\pi S)} \qquad (4)$$

where f is the shape factor and S is the surface area of the conductor. Shape factor for ellipsoid changes from 1 to 0.9, i.e. from sphere to a spherical disc. For a sphere the C is given by $\varepsilon_0\varepsilon (4\pi R)$ [15], where R is radius of the sphere. Capacitance of

non- spherical molecular conductors can be calculated using a sphere with equivalent volume [13]. In this case

$E_b = e^2 / (4\pi \varepsilon \varepsilon_0 R)$   (5)

When $E_b$ is plotted verses $e^2/ (4\pi \varepsilon \varepsilon_0 R)$, (where R is the radius of sphere with equivalent volume of the molcule). (Figure 2) we see that $E_b$ varies proportionally with ($e^2/ (4\pi \varepsilon \varepsilon_0 R)$. However it deviates from straight line with slope 1 and intercept zero.

The planner conjugated molecules can be viewed as ellipsoids with the contour of electron clouds on atomic frame work of molecule (see figure 3 inset). From equation 4 we know the capacitance varies with $f \sqrt{S}$. For equivalent volume, the surface area of ellipsoid is higher than sphere. Deviation from spherical shape will increase the capacitance and hence the less is the $E_b$. The assumption of spherical space of molecules though predicts the trend, deviates from expected correlation (from slope one and intercept zero), as the molecules deviates more from spherical shape. If we focus on polyacenes where the ellipsoidal molecular shape can be considered with semiaxial dimensions with a, b and c. Here a, b remains nearly constant with and c increases as number of rings increases (see supporting Info). In such a case "R" in equation 5 can be replaced with wR, where "w" is defined

$$w = f(\sqrt{S_{ellipsoid}}/\sqrt{S_{sphere}})$$

Figure 3 shows plot of $E_b$ versus $e^2/ (4\pi\varepsilon\varepsilon_0 wR)$. (see supporting info. for calculation of w). We can see correlation is with slope 1.07 and intercept of 0 (with regression coefficient = 0.99). With shape factor correction polyacenes show the predicted correlation. The calculation of $E_b$ using "R" does not warrant a shape factor correction if the shape of concerned molecule does not deviates drastically from a spherical shape. The value of w is from 1-1.3 in these cases.

In an ordered solid, the exciton binding energy will be less than isolated molecule in a dielectric medium, due to intermolecular interaction and band formation. However due to small intermolecular interactions the bandwidth of organic molecules are few tens of meV . This simple model only gives a first order calculation of $E_b$ due to dielectric and size of the molecule, hence should be used as near estimation (or upper limit) of exciton binding energy for well ordered crystalline material.

*Transport gap, optical gap and exciton binding energy.*

Transport and optical gap in these systems are reflection of extend of conjugation and dielectric constant of these material. As the molecular size increases, the electron and hole are delocalized over the whole molecule and reflected in lowering in the optical gap and transport gap. $E_t$ decreases much faster than $E_{opt}$ as a result $E_b$ decreases with decrease in $E_t$ and $E_{opt}$. Figure 4 shows the decrease $E_b$ in with respect to $E_t$ and $E_{opt}$. This phenomena is akin to inorganic counterpart where $E_b$ decreses with decrease in bandgap. We would like to note that in organic crystals, the exciton may delocalize over more than one molecule and in that case the exciton binding energy would be lower than predicted by this simple model.

## 4. Conclusion:

In conclusion, we show that for planar molecules the exciton binding energy can be calculated treating molecules as conductor, and is given by a simple relation BE ≈ $e^2/(4\pi\varepsilon_0\varepsilon R)$, where ε is the dielectric constant and R is the equivalent radius of the molecule. However if the molecule deviates from spherical shape, a minor correction factor should be added.

**Acknowledgement:**

Part of this work was supported by Tata Institute of Fundamental Research (TIFR), India. The author wish to thank Prof. N. Periasamy (TIFR, India), Prof. K. L. Narasimhan (TIFR, India), Prof. D. Cahen (WIS, Israel) and Prof. G. Hodes (WIS, Israel) for their invaluable suggestions.

**Table 1**

Calculated values for molecular volume, dielectric constant, IP, EA, Et, $E_{opt}$ and $E_b$. Experimental values for molecular volume and $E_{opt}$ (in thinflm, otherwise mentioned) for some molecules are given for comparison.

| Molecule | Mol. volume (Å)$^3$ | Mol. volume From density (Å)$^3$ | "R" (Å) | ε | IP (eV) | EA (eV) | $E_t$ (eV) | $^b E_{opt}$ (eV) | Exp. $E_{opt}$ (eV) | Ref. For Exp. $E_{opt}$ | $E_b$ (eV) | $e^2/(4\pi\varepsilon_0\varepsilon R)$ (eV) |
|---|---|---|---|---|---|---|---|---|---|---|---|---|
| **Napthalene** | 189.72 | 183.11[a] | 3.564 | 2.43 | 6.77 | 0.80 | 5.97 | 4.30 | 4.16[c] | [16] | 1.67 | 1.66 |
| **Anthracene** | 243.50 | 236.71[a] | 3.874 | 2.95 | 5.97 | 1.62 | 4.35 | 3.19 | 3.13 | [17] | 1.16 | 1.26 |
| **Tetracene** | 290.30 | 280.80[a] | 4.108 | 3.69 | 5.4 | 2.2 | 3.2 | 2.39 | 2.38 | [18] | 0.81 | 0.95 |
| **Pentacene** | 331.00 | 347.47[a] | 4.291 | 4.82 | 5.03 | 2.63 | 2.4 | 1.85 | 1.85 | [19] | 0.55 | 0.70 |
| **Hexacene** | 433.27 | 406.90[a] | 4.694 | 4.71 | 4.83 | 2.86 | 1.97 | 1.46 | 1.6[d] | [20] | 0.51 | 0.65 |
| **Pyrene** | 247.88 | 264.18[a] | 3.897 | 3.88 | 5.91 | 1.61 | 4.30 | 3.57 | 3.56[c] | [16] | 0.73 | 0.95 |
| **Chrysene** | 292.84 | 297.49[a] | 4.119 | 3.20 | 6.19 | 1.34 | 4.85 | 3.70 | 3.4 | [21] | 1.15 | 1.09 |
| **Perylene** | 308.80 | 310.29[a] | 4.193 | 3.45 | 5.56 | 2.05 | 3.51 | 2.68 | 2.75[c] | [16] | 0.83 | 1.00 |
| **Ntcda** | 258.50 | NA | 3.952 | 2.89 | 8.26 | 3.75 | 4.51 | 3.28 | 3.3 | [22] | 1.23 | 1.26 |
| **Ptcda** | 363.90 | NA | 4.429 | 4.63 | 6.62 | 3.86 | 2.76 | 2.26 | 2.25 | [23] | 0.50 | 0.70 |
| **mPtcdi** | 498.74 | NA | 4.919 | 3.52 | 6.44 | 3.5 | 2.94 | 2.36 | ~2.3 | [24] | 0.58 | 0.83 |

[a] Density (g/cm$^3$): Naphthalene: 1.162; Anthracene: 1.25; Tetracene: 1.35; pyrene: 1.271; Chrysene: 1.274 and perylene: 1.35 (from ref [25]); pentacene: 1.33; hexacene: 1.35.from ref [26])

[b]. Optical band gap is calculated using $n^2 = 2$ for all molecules, where n is the index of refraction

[c] Determined from absorption spectrum recorded in cyclohexane.

[d] Determined from absorption spectrum recorded in 1,2,4-trichlorobenzene.

Figure 1

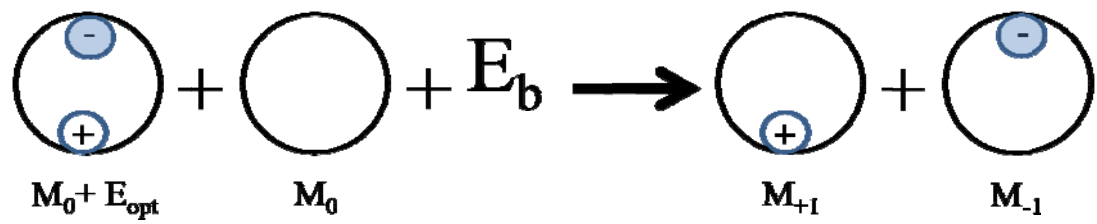

Figure1: Schematic diagram of exciton breaking.

Figure 2

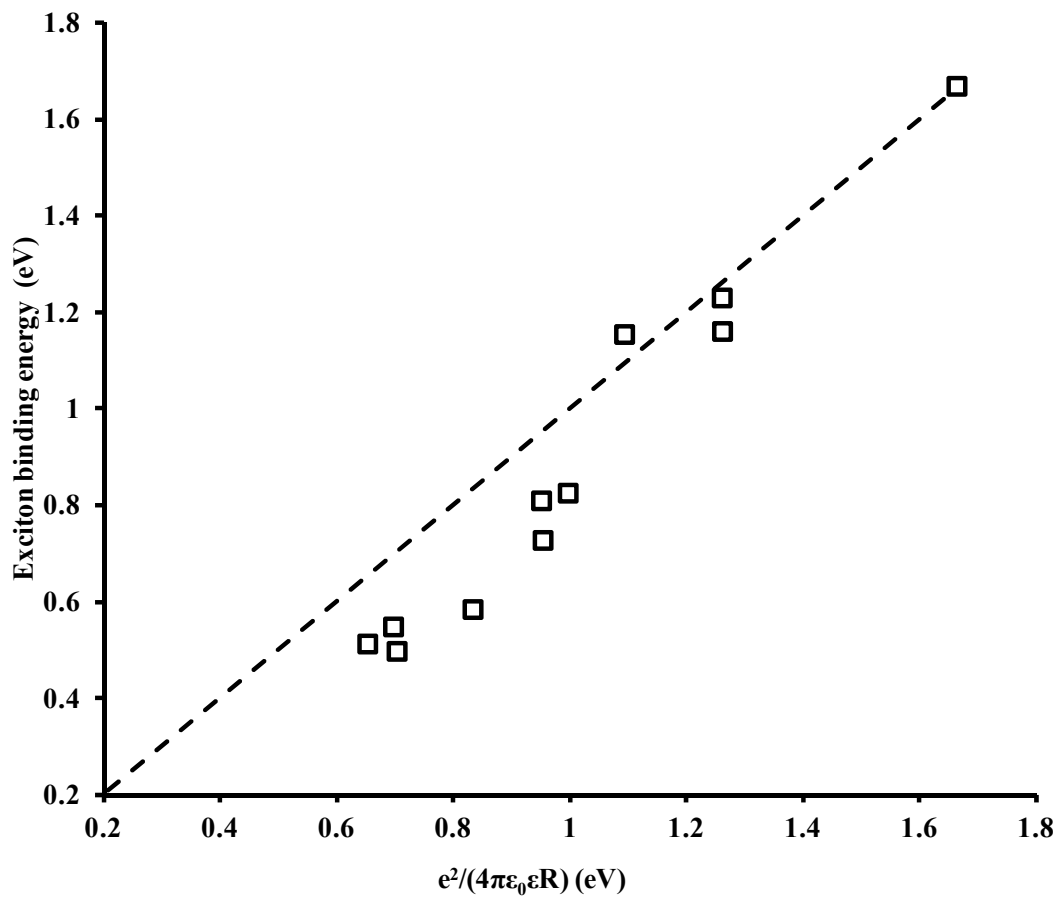

Figure 2: Plot of exciton binding energy versus charging energy of molecular conductor. The dotted line represents straight line with slope1 and intercept zero.

Figure 3

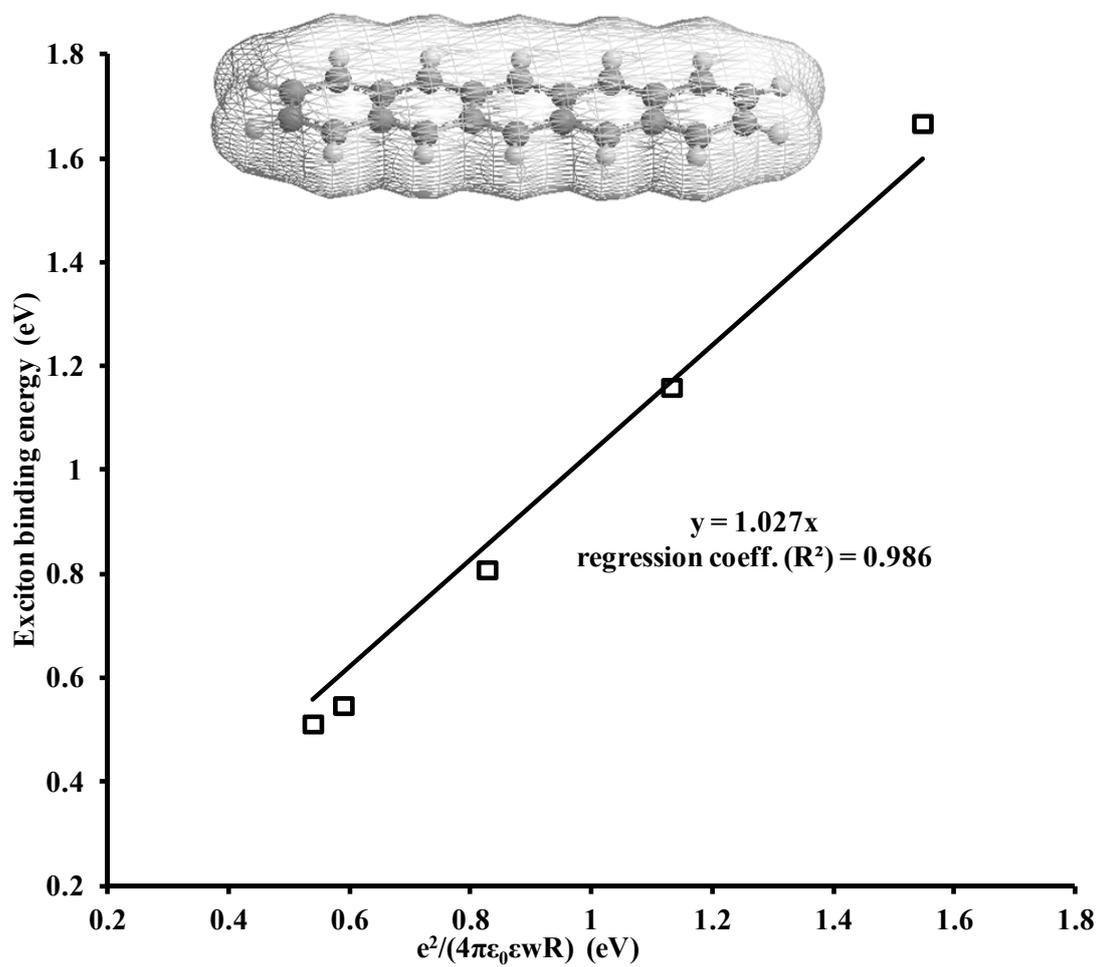

Figure 3: Exciton binding energy of polyacenes versus charging energy (with shape correction) of molecular conductor . Inset shows pentacene with charge density surface.

Figure 4.

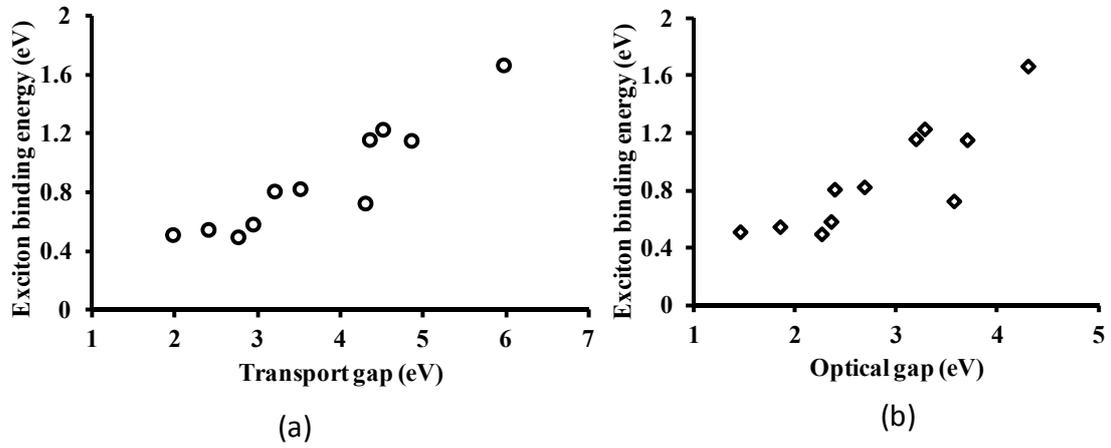

Figure 4: Exciton binding energy versus (a) Transport gap and (b) Opticalgap

**Supporting Information**

Excition binding energy in Organic semiconductor .

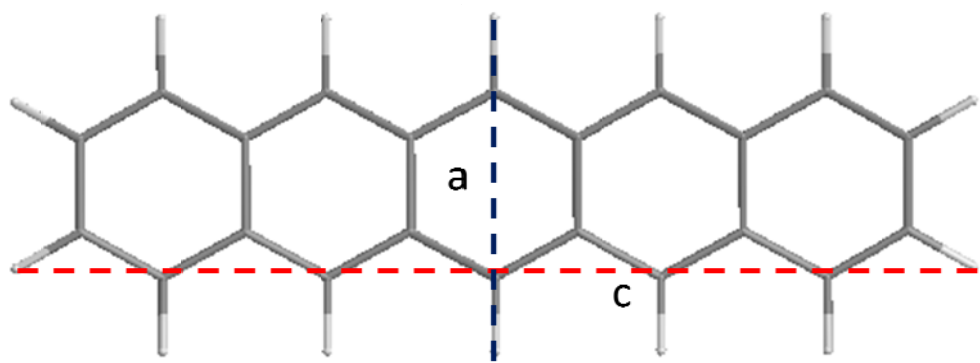

Figure S1: Here a representative molecule pentacene shows the semiaxial dimensions a, and c, the other dimension b = (3×Molecularvolume)/(4πac)

Table S1 shows values used to calculate the shape correction factor w.

Where w = $f(\sqrt{S_{ellipsoid}}/\sqrt{S_{Sphere}})$ , S is the surface area .

Surface area of ellipsoid $\approx 4\pi(\frac{a^p b^p + a^p c^p + b^p c^p}{3})^{1/p}$ where p = 1.6075

Value of f is taken 0.9 due to c/b ratio ≤ 0.2.

Table S1

| Molecule | a (Å) | b (Å) | c (Å) | $S_{ellipsoid}$ (Å)$^2$ | R ( eqv.) (Å) | $S_{sphere}$ (Å)$^2$ | f | W |
|---|---|---|---|---|---|---|---|---|
| Naphthalene | 4.98 | 1.35 | 6.76 | 238.7991 | 3.57 | 159.9517 | 0.9 | 1.099675 |
| Anthracene | 4.98 | 1.27 | 9.21 | 317.9452 | 3.88 | 188.8493 | 0.9 | 1.16778 |
| Tetracene | 4.98 | 1.19 | 11.67 | 396.9162 | 4.10 | 211.7226 | 0.9 | 1.232277 |
| Pentacene | 4.98 | 1.12 | 14.13 | 475.8013 | 4.29 | 230.9756 | 0.9 | 1.291732 |
| Hexacene | 4.98 | 1.25 | 16.59 | 563.6217 | 4.69 | 276.5573 | 0.9 | 1.284825 |